\documentclass[a4paper,12pt]{article}
\usepackage{a4wide}

\newcommand{\matrice}[4]{\left(\begin{array}{cc} {#1} & {#2} \\ {#3} & {#4} \end{array} \right) }

\def\beq{\begin{eqnarray}}    
\def\enq{\end{eqnarray}}      

\def\R{{\hbox{{\rm I}\kern-.2em\hbox{\rm R}}}}   
\def\H{{\hbox{{\rm I}\kern-.2em\hbox{\rm H}}}}   
\def\N{{\hbox{{\rm I}\kern-.2em\hbox{\rm N}}}}   
\def\C{{\ \hbox{{\rm I}\kern-.6em\hbox{\bf C}}}} 
\def\Z{{\hbox{{\rm Z}\kern-.4em\hbox{\rm Z}}}}   

\begin{document}
\begin{center}

{\LARGE \bf New physics in the} \\ 
{\LARGE \bf charged relativistic Bose gas} \\
{\LARGE \bf using zeta-function regularization?\footnote{Talk given at the ``SEWM2000 - Strong and
Electro-Weak Matter 2000'' Conference, June 2000,
Marseilles, France. Proceedings to be published by World Scientific.}}
\\[4mm]
{\large Antonio Filippi}
\\[4mm]
{\it Theoretical Physics Group, Imperial College, \\ Prince Consort
Road, London SW7 2BZ, United Kingdom\\E-mail: a.filippi@ic.ac.uk}
\\[4mm]
\end{center}

\abstract{The multiplicative anomaly, recently introduced in QFT,
plays a fundamental role in solving some mathematical inconsistencies of the widely
used zeta-function regularization method. Its physical relevance is
still an open question and is here analyzed in the light of a
non-perturbative method. Even in this approach the ``different
physics'' seems to hold and not to be easily removable by renormalization.}

\section{Introduction}

The evaluation of functional determinants of pseudo-differential
operators is often a fundamental point in quantum field theory
calculations.
As these operators are unbounded the corresponding determinant is
undefined, unless a rigorous regularization procedure is applied.
One of the most successful is the zeta-function regularization method \cite{dowcri}-\cite{bcvz}.
This uses the so called ``generalized Riemann zeta function''  $\zeta(s|A)=\mbox{Tr}\, A^{-s}$,
(A a pseudo-differential operator) which is well defined for a sufficiently large real part of $s$ and can be analytically
continued to a function meromorphic in all the plane and analytic at
$s=0$. As such its derivative with respect to $s$ at zero is well defined and
the logarithm of the zeta-function regularized functional determinant
will then be defined by
\begin{eqnarray}
\ln\det\frac{A}{\sigma^2} =-\zeta'(0|A)-\zeta(0|A)\ln \sigma^2 ,
\end{eqnarray}
where $\sigma^2$ is a renormalization scale mass.

The recent introduction \cite{evz}-\cite{regensburg} of the
long-overlooked 
multiplicative anomaly \cite{wod}-\cite{konvis}
in this framework has provoked a lively debate
\cite{byt}-\cite{mcktom2}.
Its mathematical rigor and importance for
quantum field theory has been widely
proved by this and other authors \cite{mcktom} but the question of its physical
relevance is still open.

The multiplicative anomaly can appear when we evaluate a quantity of the
form $\ln \det (A B)$, with $A$ and $B$ two commuting pseudo-differential operators.
The fact is that it is not always true that the equality $\ln \det
(AB)=\ln \det(A)+\ln \det (B)$ holds, as normally assumed in
quantum field theory until recently.
On the contrary, using zeta-function regularization, an additional term $a(A,B)$,
called the `multiplicative anomaly,
may be present on the right hand-side. The anomaly can rarely be computed
directly as the difference of the two sides. Fortunately Wodzicki.
produced a very useful indirect formula for its evaluation (see\cite{wod}).

The fact that the anomaly can add an
``anomalous'' term to the ``standard'' result, the controversial physical relevance
of which is still under scrutiny, will be the subject of this work.

For some of the very simple models studied up to now
the multiplicative anomaly is not always present,
while in others its ``anomalous term'' can be easily renormalized away
at one-loop
by a finite shift in the counterterms \cite{evz}.
Renormalizability is therefore a fundamental issue to analyze to
understand the physical content of the multiplicative anomaly.

To this end I will need to go beyond the one-loop approximation
used until now.
Since we finally want to analyze the symmetry
breaking transition, the
appropriate candidate seems a self-consistent approach similar to the
large $N$ expansion \cite{raybook}.

The relativistic charged Bose gas at finite temperature has long been of
interest on its own\cite{kap1}-\cite{chemical}. The large N limit
of the $0(N)$ interacting field can be found in Haber and
Weldon. None of these remarkable papers adopts a fully regularized
approach as there are
formal manipulations of infinite sums involved.
We showed how the anomaly is crucial for
the consistency of the zeta-function method and how it creates an
unexpected ``anomalous term''. I refer the reader to the relevant
papers \cite{efvz}-\cite{regensburg} for details and results.

\section{The non-perturbative approach}

The inclusion of the chemical potential $\mu$ in a Hamiltonian representation
of the grand-canonical partition function is usually (see \cite{kap1,habwel3,bbd,chemical})
realized by defining an effective Hamiltonian $H=H_o + \mu Q$ where $Q$ is the
charge density operator. Through integration on the momenta the
functional integral can be cast in Lagrangian form.
Although I am
only interested in $N=2$ I will leave the $N$ in the interacting term
for clarity, its sum over
repeated indices $a=1,2$ understood. The starting action is therefore
\begin{equation}
S[\phi]\!\!=\!\!\int_0^\beta\!\!\! d\tau\int\!\!\! d^3 x
\!\left(\partial_{\nu}\phi_{a}\partial^{\nu}\phi_{a}\!\!+\!m^2_{o}\phi^2\!\!+\!\frac{\lambda_{o}}{4N}\phi^4\!\!-\!\mu^2\phi^2\!\!+\!2i\mu(\phi_{2}\partial_{\tau}\phi_{1}\!\!-\!\phi_{1}\partial_{\tau}\phi_{2})\right)
\end{equation}
I can then insert in the functional integral a Gaussian representation
of unity, through an auxiliary field $B(x)$ \cite{raybook}.
Considering constant sources $J_a$ for the $\phi_{a}$ fields,
 the partition function then  becomes
\begin{equation}
Z[J]=\int{\cal D}B{\cal D}\phi_{1}{\cal D}\phi_{2}e^{-\frac{1}{2} \int dx [\phi_{a}(x)A_{ab}(x)\phi_{b}(x)-\frac{N}{\lambda_{o}}B^2(x)]}
\end{equation}
where I defined the matrix valued differential operator
\begin{equation}
A(x)=\matrice{-\partial^{2}_{\tau}-\nabla^2+m^2_{o}+B(x)-\mu^{2}}{-2i\mu\partial_{\tau}}{2i\mu\partial_{\tau}}{-\partial^{2}_{\tau}-\nabla^2+m^2_{o}+B(x)-\mu^{2}} \label{aop}
\end{equation}
 The functional integral in $\phi_a$ can be recast in exponential form to
contribute in an effective action for the $B(x)$ field.
It is also clear that this effective action is of order $N$.
We can therefore apply a saddle point approximation in the field $B(x)$,
such that

\begin{equation}
Z[J]\simeq\int{\cal D}\phi_{1}{\cal D}\phi_{2}e^{-\frac{1}{2} S[\phi,\bar{B}]+ \frac{1}{\beta}J_{a} \int dx \phi_{a}(x)}
\end{equation}
where constant $\bar{B}$ is the large-N saddle-point.

At this point it proves useful to turn to the momentum
representation,
\begin{equation}
Z[J]=e^{\frac{\beta V N}{2\lambda_o}\bar{B}^2+\frac{V}{2\beta}J_{a}{\cal
A}_{0ab}^{-1}J_{b}}e^{-\frac{1}{2}\log\det(A\sigma^2)}
\end{equation}
where I denote the eigenvalues matrix  as ${\cal A}(n,k)$ and  ${\cal
A}(0,0)$ as ${\cal A}_0$. Then
\begin{equation}
W=-\frac{1}{\beta V}\log Z = -\frac{N}{2\lambda_o}\bar{B}^2-\frac{1}{2\beta^2}J_{a}{\cal
A}_{0ab}^{-1}J_{b}+\frac{1}{2\beta V}\log\det(A\sigma^2)
\end{equation}
where ${\bar B}$ is
\begin{equation}
\frac{ N}{\lambda_o}\bar{B}\!+\!\frac{1}{\beta V}\frac{\partial}{\partial
\bar{B}}\left(\!-\!\frac{1}{2}\log\det(A\sigma^2)\right)\!-\!\frac{1}{2\beta^2
}\left(J_{1}({\cal A}_{011}^{-1})^2 J_{1}\!+\!J_{2}({\cal A}_{022}^{-1})^2 J_{2}\right) \!\!=\!\! 0
\end{equation}

Going back to (\ref{aop}) we can see that $\bar{B}$ gives a
contribution to the mass of the field $\phi$, $ M^2=m_o^2 + \bar{B}
\label{effmass} $.
With this substitution the operator $A$ has
the same form as the one for the non-interacting field.
Calculations are then highly simplified and we can resort to previous
results in \cite{efvz}-\cite{regensburg}, for which zeta-function 
regularization was used and in which the multiplicative
anomaly contribution is properly accounted for:
\begin{equation}
-\!\frac{1}{2}\log\det(A\sigma^2)\!\!=\!\!-\!\frac{\beta
V}{32\pi^2}M^4\left(\log(M^2\sigma^2)\!-\!\frac{3}{2}\right)\!+\!S
\underbrace{\!-\!\frac{\beta
V}{16\pi^2}\mu^2\left(M^2\!-\!\frac{\mu^2}{3}\right)} \label{vecchio}
\end{equation}
where $S$ is the standard particles-antiparticles thermal term,
and the underlined part is the one resulting from the multiplicative anomaly.

The constant semi-classical fields are 
\begin{equation}
\bar{\phi}_1=\beta\frac{\partial W}{\partial
J_1}=-\frac{1}{\beta}{\cal A}_{011}^{-1} J_{1}\ \ \ \ \ \ \ \ \ \bar{\phi}_2=\beta\frac{\partial W}{\partial J_2}=-\frac{1}{\beta}{\cal A}_{022}^{-1} J_{2}
\label{semiclfield}
\end{equation} 
so that, using (\ref{vecchio}), we can express $M^2$ in terms of $\bar{\phi^2}$ as
\begin{equation}
M^2 \!\!\!=\!\!\! m_o^2\!+\!\frac{1}{16\pi^2}\frac{\lambda_o}{N}M^2\log(M^2\sigma^2)-\!\frac{\lambda_o}{N\beta
V}\frac{\partial S}{\partial M^2}\!+\!\frac{\lambda_o}{2N}\bar{\phi}_a\bar{\phi}_a
\!+\underbrace{\!\frac{1}{16\pi^2}\frac{\lambda_o}{N}\mu^2} \label{mquadrin}
\end{equation}
and
\beq  
\Gamma[\bar{\phi}]&=&M^2\left[\frac{N}{\lambda_o}m_o^2+\frac{1}{2}\bar{\phi}^2\right]+M^4\left[\frac{1}{32\pi^2}\log(M^2\sigma^2)-\frac{1}{64\pi^2}-\frac{N}{2\lambda_o}\right] \\ \nonumber 
&-&\frac{1}{\beta V}S -\frac{1}{2}\mu^2\bar{\phi}^2-\frac{N}{2\lambda_o}m^4 \underbrace{+\frac{1}{16\pi^2}\mu^2\left(M^2-\frac{\mu^2}{3}\right)}
\enq
where we have redefined the
mass scale $\sigma^2\rightarrow\sigma^2 e$
so as to bring our notation into correspondence with that of Haber and Weldon.
It is now time to put $N=2$ for good.
Performing standard renormalization techniques it seems
that the anomaly cannot be renormalised away. This aspect needs to be
further analyzed, and details will be given elsewhere.

Explicit calculation shows that  the unbroken phase charge density is
\begin{equation}
\rho=-\underbrace{\frac{\mu
M^2}{8\pi^2}+\frac{\mu^3}{12\pi^2}}+\frac{1}{\beta V}\frac{\partial
S}{\partial\mu}
\end{equation}

For the broken phase where, after the
phase transition, $\mu=M$,  we can find the expression for
$\bar{\phi}^2$ and the charge density takes the form
\beq
\rho&\!=\!&\!\mu\frac{4}{\lambda_o}\left[\mu^2\!-\!
m_o^2\!-\!\frac{\lambda_o}{32\pi^2}\mu^2\log(\mu^2\sigma^2)\!+\!\frac{\lambda_o}{2\beta
V}\left.\frac{\partial S}{\partial M^2}\right|_{M^2=\mu^2}\!\!\!
\!-\!\underbrace{\!\frac{\lambda_o}{32\pi^2}\mu^2}\right] \\ \nonumber
&+&\frac{1}{\beta V} \left.\frac{\partial S}{\partial \mu}\right|_{M^2=\mu^2}-\underbrace{\frac{1}{24\pi^2}\mu^3}
\enq
It seems like the anomaly could change the ``standard'' transition
temperature and, for fixed charge $\rho$, we can see a 
difference in physical observables (e.g. pressure) in the two phases.

\section{Comments}

The relevance of zeta-function regularization in QFT can not be easily
dismissed. It is as reliable regularization technique as others.
The multiplicative anomaly is
indisputably necessary for mathematical consistency. These results
seem to show that the ``anomalous term'' it creates in certain
conditions is not trivially removable and could play a role in the
physical measurables of the system. 
To my knowledge this is the first extension of the zeta-function
regularization method in a non-perturbative calculation, within a
neat procedure that could be useful on other models as well.
It is clearly a non-trivial problem, and also
very interesting as it goes to the roots of one of the most used
regularization methods and of the functional integral approach itself.

\section*{Acknowledgments}
This work has been developed in collaboration with R. Rivers, Imperial
College. My thanks also go to E. Elizalde, L. Vanzo, S. Zerbini and T. Evans
for stimulating discussions.
The author wishes to acknowledge financial support from the European Commission
under TMR contract N. ERBFMBICT972020 and
the Foundation Blanceflor Boncompagni-Ludovisi, n\'ee Bildt.

\end{document}